# Thermal chiral anomaly in the magnetic-field induced ideal Weyl phase of $Bi_{1-x}Sb_x$ topological insulators

Dung Vu (1), Wenjuan Zhang (2), Cüneyt Şahin (3,4), Michael E. Flatté (3,4), Nandini Trivedi (2), Joseph P. Heremans (1,2,5)

1. Department of Mechanical and Aerospace Engineering, The Ohio State University, Columbus, Ohio 43210

2. Department of Physics, The Ohio State University, Columbus, Ohio 43210

3. Department of Physics and Astronomy, The University of Iowa, Iowa City, Iowa 52242

4. Pritzker School of Molecular Engineering, University of Chicago, Chicago, Illinois 60637

5. Department of Material Science and Engineering, The Ohio State University, Columbus, Ohio 43210

**Summary**

**The chiral anomaly is the predicted break down of chiral symmetry in a Weyl semimetal with monopoles of opposite chirality when an electric field parallel to a magnetic field is applied. It occurs because of charge pumping from a positive chirality to a negative chirality monopole. Experimental observation of this fundamental effect has been plagued by concerns about the pathways of the current. Here, we unambiguously demonstrate the thermal analog of the chiral anomaly in topological insulator bismuth-antimony alloys driven into an ideal Weyl semimetal state by a Zeeman field, with the chemical potential pinned at the Weyl points, and in which the Fermi surface has no trivial pockets. The experimental signature is a large enhancement of the thermal conductivity in an applied magnetic field parallel to the thermal gradient that follows the Wiedemann-Franz law.**

**Introduction**

In 1929, Weyl[1] developed a general theory describing relativistic equations of motion for gravitational forces and electron dynamics. Materials whose electrons are described by this theory are Weyl semimetals (WSMs). Nielsen and Ninomiya[2] predicted an experimental signature for the existence of WSMs: the chiral anomaly. Its importance extends beyond solid-state physics: it provides a mechanism for charge-parity violation[3] and the matter/antimatter imbalance in the universe; in condensed-matter materials, the analogous quasiparticle non-conservation is connected with a change in the vacuum state, thus preserving overall electron number.

Previous experimental determination methods for the chiral anomaly are tainted. First, currently investigated WSMs[4,5,6] are not ideal; their Fermi surfaces contain features other than Weyl nodes. Second, the signature chiral anomaly feature was a negative longitudinal magnetoresistance (MR)[7]; however, the applied magnetic field distorts current lines, sometimes complicating measurement interpretation.

Ideal WSMs have two distinguishing characteristics. First, the band structure has linearly dispersing bands that intersect at Weyl points (WPs) in a system that breaks time reversal symmetry (TRS) or inversion symmetry (IS). Second, the electrochemical potential $\mu$ is at the WP energy ($\mu = 0$) within the experimental energy range. In an ideal WSM, there are no trivial bands at energy $\mu$ and no unintentional doping. Thus, $\mu$ is pinned to the WPs because those points have the minimum system density of states (DOS); an ideal WSM displays no Shubnikov-de Haas (SdH) oscillations. Then, the Fermi surface consists only of WPs with opposite Berry curvatures, $W_R$ right-handed or $W_L$ left-handed. One pair of WPs in the Brillouin zone (BZ) is the minimum required by the Nielsen-Ninomiya theorem for a TRS-breaking ideal WSM. If multiple WPs exist, they must be degenerate. Experimentally, in an ideal WSM sample, the carrier-concentration imbalance due to unintentional doping must be smaller than the intrinsic concentration. At finite temperature, a nearly equal density of intrinsic holes and electrons is excited thermally.

The chiral anomaly[2] in ideal WSMs results from applying parallel electric **E** and magnetic fields **B** along the direction of the WP separation. **B** separates the bands into Landau levels (LL), with a 2D DOS proportional to $1/\ell_B^2$, where $\ell_B \equiv \sqrt{\hbar/eB}$ is the magnetic length. A chirality ($\chi = \pm 1$) consequence is that in the extreme quantum limit (EQL), when only the last LL is populated, electrons have only one Fermi velocity per WP: W$_R$ ($\chi = +1$) has only right-moving electrons of velocity $v$; W$_L$ ($\chi = -1$) only left-movers of velocity -$v$. **E** shifts the electron momentum on the last LL by $\delta k \propto eE\tau$ ($\tau^{-1}$ is the inter-WP scattering rate). This generates right-movers by an amount $\delta n_{+1} \propto \delta k_{+1}/\ell_B^2 \propto E\tau/\ell_B^2$ and annihilates left movers by an amount $\delta n_{-1} \propto -E\tau/\ell_B^2$. This particle generation/annihilation process is the chiral anomaly, giving rise to an additional electric current proportional to $v$ and $\delta n_{+1} - \delta n_{-1}$, which is proportional to $E = |\mathbf{E}|$ and $|\mathbf{B}| = \mu_0 H$, where $\mu_0$ is the vacuum permeability, both applied along the z direction. The anomalous contribution to the electrical conductivity generated by $N_w$ degenerate pairs of WPs is:[7]

$$\sigma_{zz} = N_w \frac{e^2 v\tau}{4\pi\hbar\ell_B^2} = N_w \frac{e^3 v\tau}{4\pi\hbar^2} B_z \qquad (1).$$

The scientific community has long considered the resulting negative longitudinal MR the crucial experimental chiral anomaly signature.[4,5,7 8]

Negative MR was observed in many WSMs, e.g., XPn compounds (X=Nb, Ta; Pn=As, Sb)[4,5,6] and Dirac semimetals[8,9,10,11], but also in materials without Weyl points near $\mu$, e.g., XPn$_2$ compounds[12,13,14,15] and elemental semimetal Bi[8]. Broadly observing this effect revealed that the negative MR is unlikely a unique chiral anomaly signature; other, classical effects might be present, as discussed next.[16,17]

The classical effects that make longitudinal MR measurements ambiguous arise because the Lorentz force distorts the current flow spatial distribution in samples with high-mobility ($\mu_B$) electrons under a magnetic field (i.e., when $\mu_B|\mathbf{B}| > 1$). This causes extrinsic, geometry-dependent MR mechanisms. The first is *current jetting*,[5,12,17,18,19] arising in 4-contact measurements. With **B**//**E**, the Lorentz force concentrates the current in a cyclotron motion near the sample center. Progressively less current passes near the voltage probes as **B** increases, lowering the measured voltage and possibly leading to the erroneous conclusion that the resistivity decreases with **B**. The second is an extrinsic *positive geometrical MR* that arises if **B** is slightly misaligned vis-à-vis the current flow lines. In the present samples, striations present on the surface of a Czochralski-grown crystal[20] can overwhelm the MR measurements, and extreme care needs to be taken with the sample alignment and geometry (see SM). Samples with reduced cross-section and smooth edges minimize both effects. Thermal conductivity $\kappa_{zz}(H_z)$ measurements avoid problems with extrinsic MR because there is no external current flow and the lattice contribution to $\kappa$ maintains a more **B**-independent heat flux than charge flux distribution in the sample (there is a magnetic-field effect on anharmonic phonon scattering[21], but it is an order of magnitude smaller than the effects discussed here).

Energy transport in WSMs poses theoretical challenges not encountered in charge transport. From the equations of motion of charge carriers at the WP and the Boltzmann transport equation, we write the imbalance between left and right moving particles ($\delta n_\chi$) and energy ($\delta \varepsilon_\chi$, the *thermal chiral anomaly*) in the presence of both an electric field **E** and thermal gradient $\nabla_r T$ as:[22]

$$\delta n_\chi = \frac{\chi e^2 \tau}{4\pi^2 \hbar^2}[\mathbf{B}\cdot\mathbf{E}]C_0 + \frac{\chi e \tau}{4\pi^2 \hbar^2}\left[\mathbf{B}\cdot\frac{-\nabla_r T}{T}\right]C_1 \qquad (2)$$

$$\delta \varepsilon_\chi = \frac{\chi e^2 \tau}{4\pi^2 \hbar^2}[\mathbf{B}\cdot\mathbf{E}](\mu C_0 + C_1) + \frac{\chi e \tau}{4\pi^2 \hbar^2}\left[\mathbf{B}\cdot\frac{-\nabla_r T}{T}\right](\mu C_1 + C_2) \qquad (3)$$

where $C_m = \int (\varepsilon - \mu)^m \left(-\partial f_0 / \partial \varepsilon\right) d\varepsilon$, $m \in \{0,1,2...\}$ with $f_0$ the Fermi-Dirac distribution function. The thermal chiral anomaly thus has two terms: First, a temperature gradient $\nabla_r T$ alone, disregarding any induced electric field, creates an imbalance between the energy carried by the left and right movers while maintaining equal populations when $\mu = 0$ ($C_1 = 0$, $\delta n_\chi = 0$, $\delta \varepsilon_\chi \neq 0$). This response contrasts with the electrical case where $\nabla_r T = 0$ and **E** create an imbalance between the populations of left and right movers while, when $\mu = 0$, maintaining the same total energy ($\delta n_\chi \neq 0$, $\delta \varepsilon_\chi = 0$). Second, when the sample is mounted in open-circuit conditions and no external electric field is applied, applying $\nabla_r T$ induces a Seebeck electric field $\mathbf{E}=S(-\nabla_r T)$ ($S$ is the thermopower), driving both $\delta n_\chi \neq 0$ and $\delta \varepsilon_\chi \neq 0$. This creates an additional $\kappa_{zz}(H_z)$ term, the ambipolar thermal conductivity,[23] $S^2 T \sigma$ (see SM). The total thermal conductivity becomes $\kappa_{zz} = \kappa_{zz,0} + S^2 \sigma T$, where $\kappa_{zz,0}$ denotes the energy carried directly by the charge carrier. For completeness, we mention that anomalously large quantum oscillations in the $\kappa_{zz}(H_z)$ of TaAs[24] is interpreted as a manifestation of chiral zero sound. We see no evidence for this behavior in $Bi_{1-x}Sb_x$ (x>10%) alloys.

The experimental tests for these theories are first to observe an increase in electronic thermal conductivity in a longitudinal magnetic field, and second to verify the Wiedemann-Franz law (WFL) in the EQL:

$$\kappa_{zz} = LT\sigma_{zz} \qquad (4),$$

with $L$ the Lorenz ratio. If each electron carries charge $e$ and entropy $k_B$, and conserves its energy during scattering, $L = L_0 = \pi^2/3 \left(k_B/e\right)^2$. The experiment consists in testing the ratio $\kappa_{zz}/T\sigma_{zz}$, which we define as $L$, against the independent variables $H_z$ and $T$, and, if $L$ is independent of these, to verify if the value equals $L_0$. In particular, a Weyl semimetal in which inelastic scattering is limited by the inter-WP scattering time $\tau$, in the quasi-classical limit at $H=0$ is expected to have $L = 7\pi^2/5 \left(k_B/e\right)^2$ but $L=L_0$ in the EQL (see SM and Ref [25]). In the presence of ambipolar conduction $L>L_0$, because $L_0$ applies only to $\kappa_{zz,0}$. Extrinsic effects result in underestimations (current jetting) or overestimations (geometrical MR) of $L$.

In this paper we report the thermal conductivity $\kappa_{zz}(H_z)$ dependence on $H_z$, and show experimentally that the chiral anomaly affects energy and charge transport similarly, i.e., $d\kappa_{zz}/dH_z > 0$, as expected from Eq. (1) and (4). We then experimentally derive values for $L$.

Previous $\kappa_{zz}(H_z)$ measurements exist: a 1% experimental increase in $\kappa_{zz}$ for GdPtBi has been reported at $H_z$ =9 T.[26] However, those samples exhibited SdH oscillations in their MR, which proves that their $\mu$ is not at the WPs. An excess $\kappa_{zz}$ also is observed in NbP,[27] dubbed a gravitational anomaly due to the formal link[28,29] between gradients $\nabla \Phi$ in the gravitational field and $\nabla_r T$. Here, we report $\kappa_{zz}(H_z)$ in magnetic-field induced ideal WSMs, $Bi_{1-x}Sb_x$ alloys with x = 11 and 15 at.%. We demonstrate that these alloy samples, topological insulators (TIs) at $|\mathbf{B}|=0$,[30] become WSMs without trivial bands in a quantizing magnetic field along the trigonal axis (z=[001]). We further identify the WP locations. In these material samples, we show their carrier concentrations are intrinsic above ~30 K, where the relevant $\kappa_{zz}$ data are collected. This makes them ideal WSMs by construction. Their $\kappa_{zz}(H_z)$ shows an electronic thermal conductivity increase by up to 300% at 9 T. Lorenz ratio $L = \kappa_{zz}/T\sigma_{zz}$ measurements show that $L \approx L_0$. Four experimental observations support relating these thermal conductivity results to the chiral anomaly:

- The effect is robust to disorder, being observed on five samples with two different compositions and mobilities at 10 to 12 K ranging from $1.9 \times 10^6$ to $2 \times 10^4$ cm$^2$V$^{-1}$s$^-$;
- The dependence of $\kappa_{zz}$ on the direction of **B**;
- The absence of the effect in samples that are similar in composition, but fall outside the range of compositions where WSMs form; and

- The temperature dependence of the observation is insensitive to phonon scattering but shows an activated behavior, with activation energy equal to the bandwidth of the Weyl bands.

**Field-induced Bi$_{1-x}$Sb$_x$ Weyl Semimetals.**

In this section, we show Bi$_{1-x}$Sb$_x$ alloys (~9 < $x$ < ≈18 at%) become ideal WSMs in a magnetic field $H_z$ above a critical threshold $H_C$ in three steps: We establish (1) their conduction and valence bands cross at $H_C$, (2) two crossing points appear that are Berry curvature monopoles, i.e., WPs, that increase further in field, and (3) there are no trivial bands.

First, to determine that the conduction and valence bands cross at $H_C$, we examine the band structure evolution of Bi$_{1-x}$Sb$_x$ alloys as a function of $x$ composition and an applied magnetic field along the [001] direction (**Figure 1**). At zero magnetic field (**Fig. 1(a)-(e)**), the alloys evolve with increasing $x$ through four successive types:[31] conventional semimetals, semimetals with an inverted band at the BZ L-point, semiconductors, and TIs. Even though the band structure (**Fig. 1(a)**[31]) is known experimentally, we needed a band-structure calculation to derive the Landé factors, which form a $g$-tensor (rather than a $g$-factor) in the Bi-Sb alloy system. The anticipated electronic structure details of the Bi-Sb alloy change slowly as band positions change relative to the chemical potential. The nearly unchanged intrinsic spin-Hall conductivity calculated through the semimetal-TI transitions indicates this.[32] A tight-binding Hamiltonian describes the band structure of unalloyed bismuth and antimony,[33] incorporating the $s$- and $p$-orbitals of the two atoms in the conventional hexagonal unit cell. The alloy electronic structure is calculated (see SM) using a modified virtual crystal approximation (VCA) in which the tight-binding parameters are obtained directly from those of the elemental semimetals. The calculated band-edge evolution shown in **Fig. 1(a)** agrees with previous experiments[34,35,36] within the experimental uncertainty on compositions (1 at.%).

With these parameters, we show that a quantizing magnetic field along the trigonal direction of the TIs (**Fig. 1(f)-(k)**) inverts the bands again. The total Hamiltonian $H = H_{tb} + H_{SOC} + H_{Zeeman}$, with $H_{Zeeman} = -g_{eff}\mu_B S \cdot H \equiv -S \cdot H$, describes the Zeeman coupling effect on the alloy electronic structure $H_{tb} + H_{SOC}$, including spin-orbit coupling. The $g_{eff}$-tensors at the high-symmetry BZ L- and T-points are calculated per Ref. [37] for valence and conduction bands. The T-point $g_{eff}$-tensor has only one non-zero component, $g_{hz} = 20.5$, which only couples to the magnetic field along $z$. The more complicated effective $g_{eff}$-tensor at the L-point shows significant asymmetry. For the conduction and valence bands at the Bi$_{89}$Sb$_{11}$ L-point with a magnetic field applied along the trigonal direction, the calculated values are $g_z = -77.5$ and $-72.3$, respectively. SdH oscillations[38] in Bi confirm the extremely large $g$-factor values experimentally. This results in an anomalously large effective Zeeman splitting energy $\Delta\varepsilon_z = -\mu_B g_z B_z \approx -4.2$ meV/T at the L-point that overwhelms the orbital splitting of the LLs. Consequently, the band gaps close (**Fig. 1f,h**) at a critical field $H_C$, calculated to be ~3 T for alloy compositions near x=11%. Magnetic-field-induced band closings are uncommon, but have been reported via magneto-optical measurements on Bi[39]. $H_C$ is sensitive to parameter values used in calculations, and is of the order of 1−4 T. At $H_z > H_C$, the Zeeman energy increase further splits the degeneracy of the Kramers doublets (points W in **Fig. 1(i)**).

For step two, we demonstrate that the Kramers doublets become WPs resulting from explicit TRS breaking by showing that the Chern number changes by an integer for a momentum slice taken between these WPs. The Chern number is an integer that counts the monopoles enclosed in a given Gaussian surface in the BZ.

We calculate the Berry curvature distribution $\Omega_n(\mathbf{k})$ in momentum space for the alloy band structure to search for the WPs where the Berry curvature is concentrated and singular (see SM). The two WPs carry monopole Berry curvature $\Omega(\mathbf{k}) = \chi \mathbf{k}/k^3$ with opposite chirality, $\chi = \pm 1$. Integrating the Berry curvature provides the Chern number (see SM). A Chern number integer change provides evidence of a topology change and existence of WPs, a pair of points separated symmetrically near each L-point in the 3D BZ (**Fig. 1(j)**). WPs are located by determining the Berry curvature monopole locations and opposite chirality nodes $\chi = \pm 1$ (position provided schematically in **Fig. 1(k)**; precise coordinates given in the SM for an external field $H_z$ = 8 T). The separation between the two WPs is in the binary-trigonal plane with a major component along the trigonal direction, which coincides with the external magnetic field direction and a minor component along the bisectrix direction.

Finally, to ascertain that the $Bi_{89}Sb_{11}$ system is an ideal WSM at $H_z > H_C$, the model verifies that no trivial bands contribute to transport: neither the T-point band nor any new bands move near $\mu$ with increasing $H_z$. In a semiconductor or semimetal without unintentional doping, $\mu$ is pinned at the energy of the lowest DOS, which, without trivial bands, occurs at the WPs. Therefore, if we can minimize unintentional doping experimentally, our experimental systems described below form ideal WSMs by construction.

**Experimental**

Evidence for the thermal chiral anomaly is shown in six single-crystal samples of $Bi_{1-x}Sb_x$, $x = \sim 11$ and 15 at%. For control, we report the absence of the anomaly in two semi-metallic samples with $x \approx 5\%$; for this composition an ideal WSM does not exist. The sample compositions and characterizations are presented in the methods section. The temperature dependence of the resistivity and low-field Hall effect of the best samples (#1 with x=11% and x=15%) are used to derive carrier concentration and mobility (**Fig. 2 (a-b)**) showing that charge carriers freeze out. This, and the absence of SdH oscillations in the high-field longitudinal magneto-resistivity down to 2 K (SM) and other transport properties (SM), indicate that they are ideal WSMs. The zero-field thermal conductivity $\kappa_{zz}$ along the trigonal direction of sample #1 is given in **Fig. 2(c)** (for x=15%, SM). It consists of a phonon $\kappa_L$ and electronic $\kappa_E$ contribution separated by measuring $\kappa_{zz}$ ($H_y$) (SM) which shows a steady decrease to a saturation value at high field. This is the ordinary behavior of high-mobility materials[40,41] used to isolate $\kappa_L = \lim_{H_y \to \infty}(\kappa_{zz}(H_y))$ for $T<120$ K. At $T>120$ K, $\kappa_L(T)$ is extrapolated following a $T^{-1/3}$ law[41] to 300 K. $\kappa_L$ dominates $\kappa_{zz}$ below 35 K, limiting measurements of $\kappa_E$ to $T>35$ K. At zero field, $\kappa_E(H_z=0)$ follows the WFL with $L=L_0$ (dashed line in **Fig. 2(c)**) above 30 K.

**Figure 3** shows the longitudinal magneto-thermal conductivity $\kappa_{zz}(H_z)$ of three samples: **(a)** $Bi_{95}Sb_5$ (not a WSM); **(b)** $Bi_{89}Sb_{11}$, and **(c)** $Bi_{85}Sb_{15}$; (both WSMs above 1−2 T). $\kappa_E(H_z)$ of $Bi_{89}Sb_{11}$ is reported as a function of $H_z$ in **Fig. 3(d)**: the relative $\kappa_E$ increase in magnetic field reaches above 300% from 34 to 85 K at 9 T. At low field, $d\kappa_{zz}/dH_z < 0$ for $H_z<1$ T at $T <50$ K and $H_z<3$ T at $T=160$ K. Here, the last LLs of the conduction and valence bands have not crossed in energy. At high field, in WSM phase, $d\kappa_{zz}/dH_z > 0$. We posit the large increase in $\kappa_{zz}(H_z)$ (**Fig. 3(b-d)**) at high field is experimental evidence for the thermal chiral anomaly. The following observations justify the thesis. First, **Fig. 3(a)** shows that $d\kappa_{zz}/dH_z < 0$ at all fields for $Bi_{95}Sb_5$, which in zero field is a conventional semimetal, not a TI, with a trivial hole pocket in its Fermi surface at the BZ T-point. In $Bi_{95}Sb_5$, the band crossing with field does not create an ideal WSM phase; if the $d\kappa_{zz}/dH_z > 0$ observation on $Bi_{89}Sb_{11}$ and $Bi_{85}Sb_{15}$ resulted from effects other than the chiral anolmaly, e.g., ionized impurity scattering,[42] known to be weak even in doped Bi,[20] it also would occur in similarly prepared $Bi_{95}Sb_5$. Second, to ascertain that a circulating current or an artifact on the sample surfaces does not induce the effect, samples of $Bi_{95}Sb_5$ and $Bi_{89}Sb_{11}$ were mounted with its top and bottom faces covered by electrically conducting Ag epoxy (see SM). We observe no effect from the added surface conducting layers. Third, the $d\kappa_{zz}/dH_z > 0$ data at high $H_z$ were reproduced on $Bi_{89}Sb_{11}$ samples 2−4 (see SM), which had a mobility of only $2 \times 10^4$ $cm^2V^{-1}s^{-1}$ at 12 K, demonstrating the robustness of the observations vis-à-vis defect scattering. Fourth, $d\kappa_{zz}/dH_z > 0$ in **Fig. 3(b-d)** is observed up to 200 K, twice the Bi Debye temperature, demonstrating the robustness of the effect to phonon scattering. For $T>200$ K, $d\kappa_{zz}/dH_z < 0$ at all fields due to thermal smearing of the carrier population between the WPs. The last argument will be presented below: the observation has only one energy scale, the Weyl band width.

Simultaneous $\kappa_{zz}(H_z)$ and MR ($\rho_{zz}(H_z)$) measurements were taken on a specially prepared $Bi_{89}Sb_{11}$ sample (#6, see methods), and are shown **Fig. 4a**. Subtracting $\kappa_L$[41] from $\kappa_{zz}(H_z)$ gives $\kappa_{zz,E}(H_z)$. The WFL is tested by plotting the product $\kappa_{zz,E}(H_z).\rho_{zz}(H_z)$ normalized to $L_0T$ in **Fig. 4b** as function of $T$ for $H_z = 5$ and 9 T. The $H_z$-dependence of the result is within the error bars, and **Fig. 4b** verifies that the WFL holds in an applied field with $L \approx L_0$. Since the material is an ideal WSM and the WSM phase is induced in EQL, the Lorenz ratio is expected to be $L_0$. The error bar increases with decreasing $T$ as $\kappa_L$ increasingly dominates $\kappa_{zz}(H_z)$ and becomes as large as the signal below 50 K. $\kappa_L$ masks the electronic contribution completely below 35 K. This knowledge allows fitting the $d\kappa_{zz}/dH_z$ (inset in **Fig. 4c**) experimental temperature dependence at $T>60$ K. Using equations (1) and (4) with $L=L_0$, $N_w=12$ to derive the thermal chiral conductivity, then taking its field derivative, we obtain:

$$d\kappa_{zz}/dH_z = \frac{\pi e v k_B^2}{\hbar^2} T \tau \qquad (5).$$

Using the calculated $v \approx 4.5 \times 10^5$ m/s (see SM), Eq. (5) can be used to derive the inter-WP scatting time $\tau(T)$, shown in **Fig. 4(c)**. Below ~ 60 K, $\tau$ of $Bi_{89}Sb_{11}$ tends asymptotically to $10^{-12}$ s and is temperature independent at a value one order of magnitude longer than the electron relaxation time in $Bi_{95}Sb_5$ at 4.2 K. This suggests a high degree of charge-transport protection. In $Bi_{85}Sb_{15}$ and at $T > 60$ K in $Bi_{89}Sb_{11}$, $\tau$ increases exponentially with $T^{-1}$, an activated behavior with activation energy of 34±2 meV for $Bi_{89}Sb_{11}$ and 15±2 meV for $Bi_{85}Sb_{15}$. These values are expected when charge carriers are thermally excited above the Weyl bandwidth limit $\tau$. The calculated band width (see SM) at 7.5 T is $E_{BW}=35$ meV for $x = 10.5$ at.% and $E_{BW} = 20$ meV at 7.5 T for $x=15.1$ at.%, the measured concentrations in the samples. Overall, the **Fig. 4(c)** data suggest that thermal smearing of the carrier population between the WPs is the main mechanism inhibiting the observed increase in $\kappa_{zz}(H_z)$, and that $E_{BW}$ is the only energy scale in the observations.

The angular dependence of the effect, $\kappa_{zz}$ ($H_\theta$), is acquired on sample #4: **Fig. 4(d)** reports the increase $\Delta\kappa_E = \kappa_{zz}(H = 9T) - \kappa_{zz,MIN}$ ($\kappa_{zz, MIN}$ is the $\kappa_{zz}$ value at minimal field). The angular dependence follows a cos($\theta$)$^n$ law with $n>4$, a much higher exponent than expected from the component of $H_\theta$ projected along $z$.

Altogether, we posit that the $d\kappa_{zz}/dH_z > 0$ observation constitutes robust experimental evidence for the thermal chiral anomaly in an ideal WSM, impervious to current-line distortions and resulting data-interpretation ambiguities that plague MR measurements. The WFL holds at zero field and in field, with $L \approx L_0$. The robustness of the results vis-à-vis defect and phonon scattering and the fact that the only energy scale is the width of the Weyl bands point to the topological origin of the data.

**Methods**

We studied four samples of nominal composition $Bi_{89}Sb_{11}$, labeled 1-4, one $Bi_{85}Sb_{15}$ and one $Bi_{95}Sb_{5}$ sample, cut from four separate single crystals. The $Bi_{85}Sb_{15}$ (sample #5) and $Bi_{89}Sb_{11}$ (samples #1 and #6) are grown in-house by the TMZ technique (see SM); the $Bi_{95}Sb_{5}$ in-house by the Bridgeman method, and $Bi_{89}Sb_{11}$ (samples #2-4) grown by Noothoven van Goor using the Czochralski method.[20] Table 1 summarizes the sample compositions and purpose. We checked by X-ray diffraction (XRD) (see SM) the compositional uniformity of the TMZ crystal centers. The $Bi_{89}Sb_{11}$ crystal composition (providing sample #1) was 10.5±0.5 at.%, and compositional uniformity was interpolated from that of the crystal to better than 0.1% across the sample size. The $Bi_{85}Sb_{15}$ crystal composition was 15.1±0.7 at.%, and the measured sample was uniform to better than 0.1% across the sample size. We measured the low-field Hall effect and resistivity (see SM) of separate TMZ crystal pieces, given in **Fig. 2 (a-b)**. The Hall effect polarity switched from n-type to p-type, indicating almost complete charge-carrier freeze-out. Because Bi and Sb are isoelectronic, achieving freeze-out does not require the exquisite stoichiometric control needed for compound semiconductors, but the starting materials required in-house zone refinement because we could not reach carrier concentrations $<10^{17}$ cm$^{-3}$ with 99.999% pure commercial materials. Three cuts (samples #2, #3, #4) of a separate Czochralski crystal piece showed a $\kappa_{zz}$ increase in field, and had electron densities and mobilities of $8\times10^{18}$ cm$^{-3}$ and 1,050 cm$^{2}$V$^{-1}$s$^{-1}$ at 300 K, which froze out to $1.4\times10^{16}$ cm$^{-3}$ and $2\times10^{4}$ cm$^{2}$V$^{-1}$s$^{-1}$ at 12 K. Van Goor determined the composition to be 12%; XRD measurements similar to those on the TMZ crystal give 11.3±0.7 at.%.

Thermal conductivity was measured along the [001] crystal direction with the steady state method in the high-vacuum (10$^{-6}$ Torr), radiation-shielded environment of a Quantum Design

Physical Property Measurement System (PPMS) sample chamber. The heat source was a resistive heater (Omega Engineering, Inc., 120 Ω strain gauge) bonded to a $Al_2O_3$ plate heat spreader. The heat sink, also an $Al_2O_3$ plate, was glued to the base of the PPMS's AC puck. The heat source and sink were bonded to the cleaved sample top and bottom using GE varnish to ensure these surfaces were short-circuit free. The thermometers were fabricated thermocouples of 25-µm-diameter copper-constantan couples. The thermometers contacted the sample at different positions along the temperature gradient with epoxy. We conducted measurements at discreet temperatures between 10 K and 300 K. The sample assembly was stabilized thermally at each discrete temperature for 30 minutes before measurement. Magneto-thermal conductivity was measured in a sweeping-down magnetic field from 9 T to -9 T in the PPMS, with sweeping rate of 5 mT/s. Controls software was programmed using LabVIEW. Longitudinal MR measurements $\rho_{zz}(H_z)$ were carried out on sample #6, which was designed with a long, thin geometry (3 × 0.4 × 0.6 mm) with voltage probe wires attached along the spine of the sample[8] to minimize current jetting. This, and the fact that the magnetic-field alignment was controlled to ~0.1°, minimizes the geometrical MR effects. Its surface was etched to smooth out the surface damage from cutting.

The thermal conductivity measurement error is dominated by the sample geometry uncertainty, of the order of 10%, and thermocouple calibration. Heat losses were calculated from the measured instrumental heat leaks, which vary with temperature but are of the order of mW/K above 200 K, much smaller than the thermal conductance of the sample. The Cu-Constantan thermocouples were calibrated in field using a thermal-conductivity measurement on a glass sample (see SM). We presumed in data treatment that they have no magnetic-field dependence. Measured, the field dependence up to 7 T was <2% down to 80 K, <5% down to 34 K, but as much as 10% at 16 K, the lowest-reported $\kappa_{zz}$ temperature. Thus, all data uncertainty reported in **Figures 3** and **4** above 40 K is better than 14%, representing the aggregate of thermocouple calibration uncertainty and heat losses.

For angular-dependent magneto-thermal conductivity measurements, an angle between the temperature gradient and the magnetic field was created with a pre-fabricated, solid copper wedge at a desired angle. The copper wedge was treated as an additional set of thermal and contact resistances in data analysis.

**Data Availability**

The datasets generated and/or analyzed during the current study are not publicly available, but are available from the corresponding author on reasonable request.

**Acknowledgements**

This work was supported by CEM, and NSF MRSEC, under grant number. DMR-2011876. The authors acknowledge useful discussions with Dr. Maria A. H. Vozmediano. Renee Ripley edited the text and contributed to the illustrations.

**Author contributions**

The experiments were designed and carried out by D.V. and J.P.H., the theory by W.Z., C. S., M. E. F., N. T., and J.P.H. All contributed to the integration between theory and experiment and in writing the manuscript.

**Additional information**

Supplementary information is available in the online version of the paper. Reprints and permissions information is available online at www.nature.com/reprints.

Correspondence and requests for materials should be addressed to J.P.H.

**Competing financial interests**

The authors declare no competing financial interests.

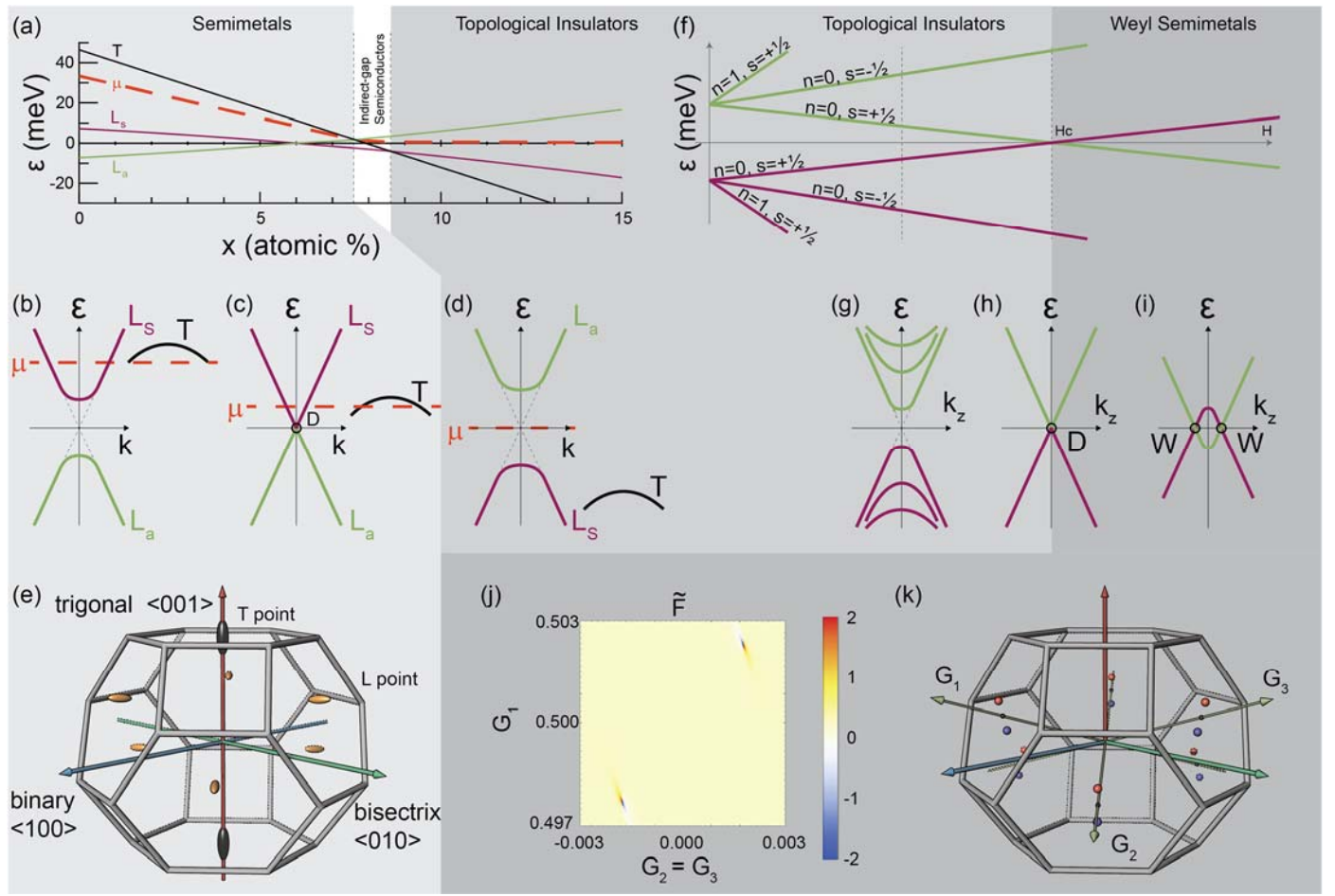

**Figure 1 Evolution of $Bi_{1-x}Sb_x$ alloys with composition and magnetic field.** (a) Band-edge energies' composition dependence at zero applied magnetic field. Elemental semimetal Bi has electrons residing in a conduction band, $L_s$, and holes in the valence band, T; with a filled second valence band, $L_a$. Adding Sb (x in at.%), the $L_a$ - $L_s$ gap closes until the bands intersect near $x \approx 6\%$. The T-band edge intersects that of the $L_a$ and $L_s$ bands at $x \approx 7.7\%$ and $x \approx 8.6\%$, respectively (from VCA, see text). The chemical potential $\mu(x)$ evolution for samples with no unintentional doping is shown as a dashed orange line. Alloys with $x < 7.7\%$ are semimetals with $\mu$ in a band; alloys with $x > 8.6\%$ are direct-gap topological insulators (TIs)[31] with $\mu$ at mid-gap in undoped material. (b) Semimetal Bi dispersion relation, (c) $Bi_{94}Sb_6$ alloys' Dirac dispersion, and (d) Bi-Sb TIs' dispersion. (e) Bi BZ and Fermi surfaces: electrons fill 6 pockets at the BZ L-points; holes fill 2 pockets at the T-points. (f) TI alloy $Bi_{89}Sb_{11}$ band-edge energies in a magnetic field $H_z$ applied along the trigonal direction. The field separates the $L_a$ and $L_s$ valence bands into Landau levels, with orbital quantum number $n$ and spin $s$. With increasing $H_z$, the $n=0$, $s=1/2$ of the $L_a$ and $L_s$ bands cross again at a critical field $H_C$. At higher fields, the crossing points develop into Weyl points (see text). (g) Dispersions along $k_z$ at $H_z < H_C$. (h) Dispersion in $k_z$ at $H_z = H_C$. (i) Dispersion at $H_z > H_C$ becomes that of a field-induced Weyl semimetal. (j) Calculated Berry curvature distribution in the **G** axes system, identified in (k). Color contrast highlights the WPs in (j); their location in the BZ is given schematically in (k), ; blue and red points indicate WPs with opposite Berry curvature.

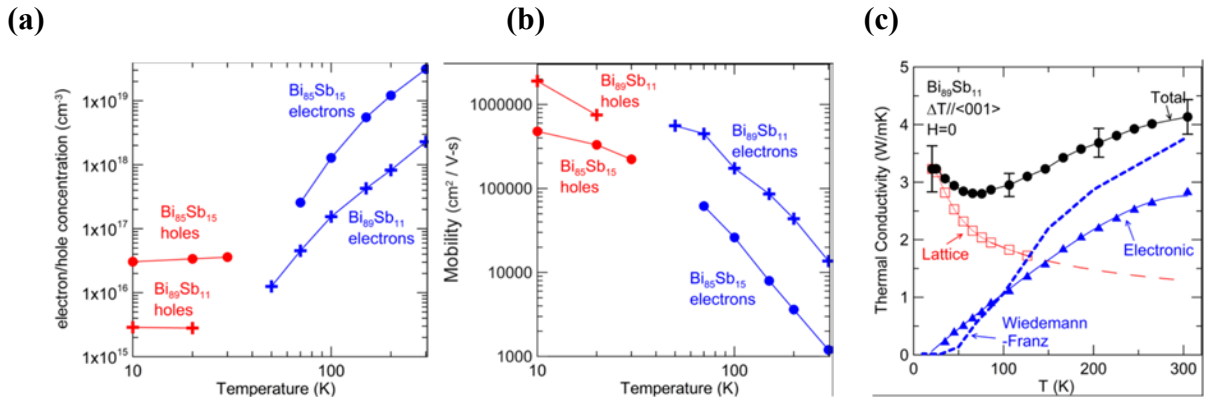

**Figure 2: Bi$_{89}$Sb$_{11}$ and Bi$_{85}$Sb$_{15}$ electronic and thermal properties .** (a) carrier concentration and (b) mobility; the samples switch from dominantly n-type at 300 K to dominantly p-type at 10 K. (c) Bi$_{89}$Sb$_{11}$ zero-field $\kappa_{zz}$ separated into lattice $\kappa_L$ and electronic $\kappa_E$ parts (see supplement for Bi$_{85}$Sb$_{15}$ data). The dashed blue line is $\kappa_E$ calculated from the resistivity and the Wiedemann-Franz law (WFL) with $L=L_0$.

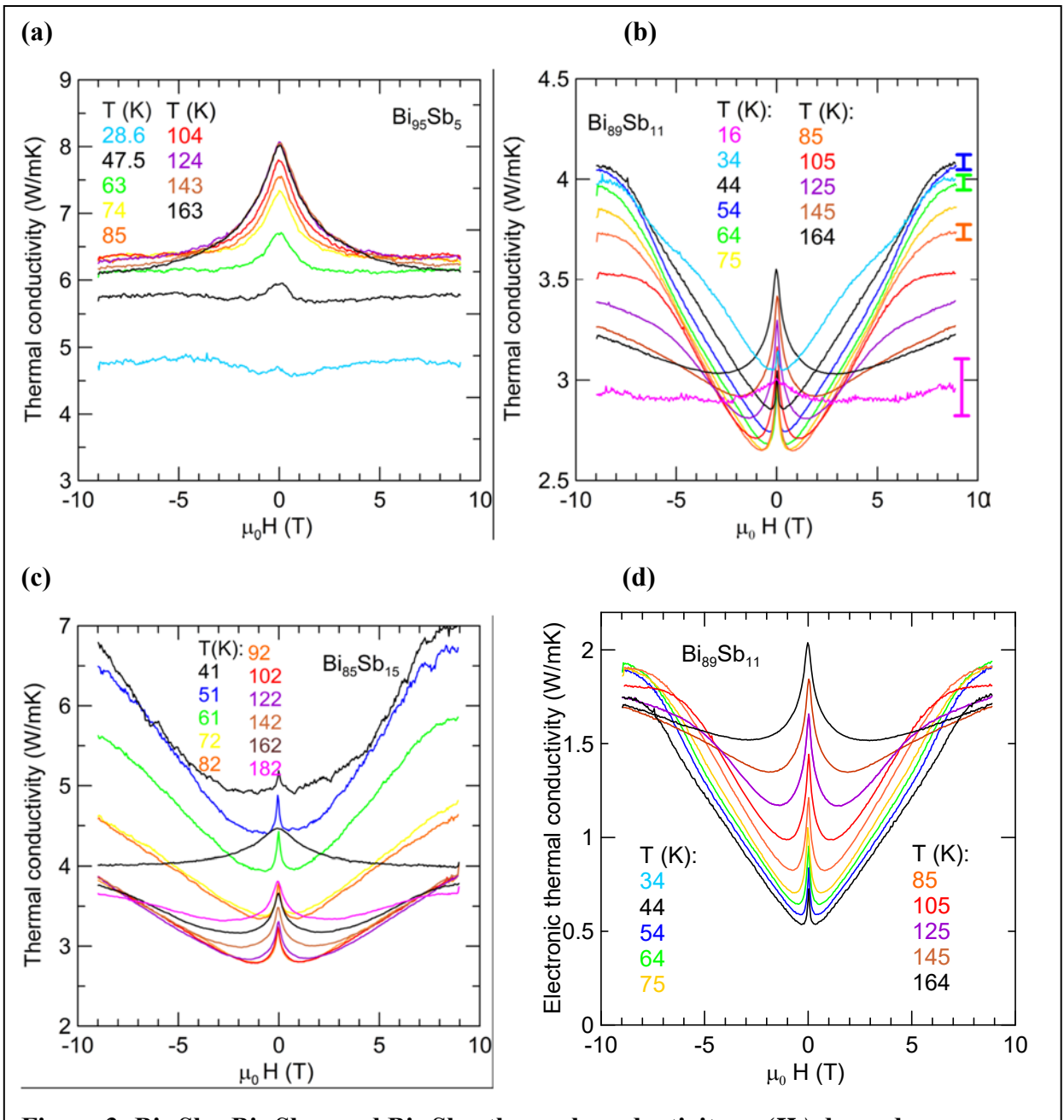

**Figure 3: Bi$_{95}$Sb$_5$, Bi$_{89}$Sb$_{11}$, and Bi$_{85}$Sb$_{15}$ thermal conductivity $\kappa_{zz}(H_z)$ dependence on longitudinal magnetic field** along the trigonal ($z$=<001>) direction at temperatures indicated. (a) Bi$_{95}$Sb$_5$, a conventional, not Weyl, semimetal, has $\kappa_{zz}$ ($H_z$) that monotonically decreases with $H_z$, due to a positive MR. The $\kappa_{zz}(H_z)$ of (b) Bi$_{89}$Sb$_{11}$ (sample 1) and (c) Bi$_{85}$Sb$_{15}$ shows a decrease due to a conventional positive MR in the TI regime, followed by an increase that we posit is evidence for the thermal chiral anomaly. The error bars (derivation in SM) are relative to the field dependence, not the absolute value. (d) The electronic contribution $\kappa_E$ of the total thermal conductivity $\kappa_{zz}$ for Bi$_{89}$Sb$_{11}$ is obtained by subtracting the lattice contribution $\kappa_L$. $\kappa_E$ shows a >300% increase with field at 9 T. The uncertainty in measurements is described in the methods section.

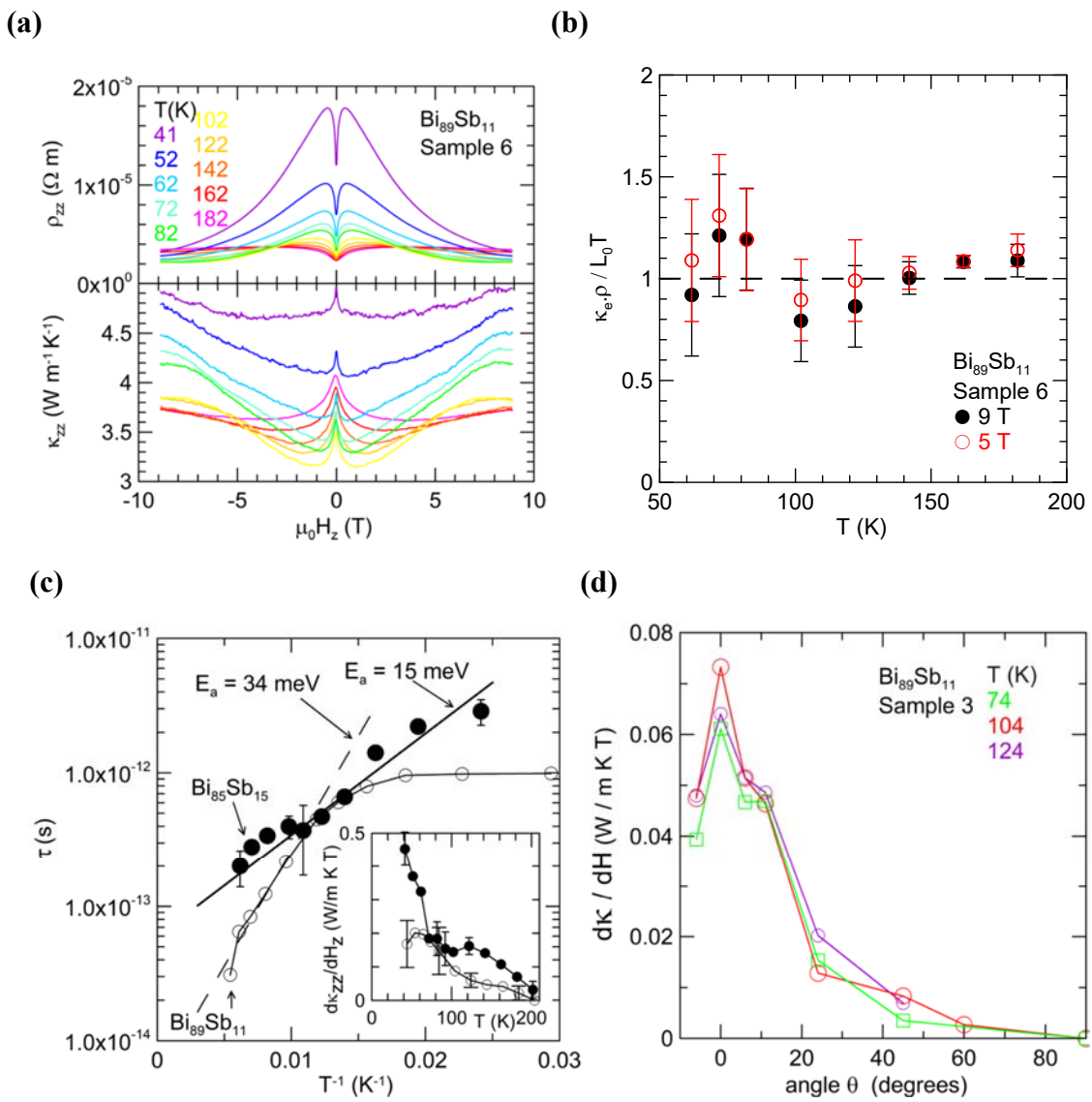

**Figure 4 Wiedemann-Franz law verification; temperature and angular dependence of the $\kappa_{zz}$ ($H_z$) increase.** (a) Bi$_{89}$Sb$_{11}$ (sample 6) $\kappa_{zz}$ ($H_z$) and $\sigma_{zz}$ ($H_z$) (b) Lorenz ratio $L = \kappa_{zz,e}$ ($H_z$) / $\sigma_{zz}$ ($H_z$) derived from (a), normalized to $L_0T$ at two values of $H_z$; $L$ is independent of $H_z$ within the error bar; (c) The inter-WP scattering time $\tau$, derived from equation (5) fits Arrhenius plots (lines) at $T > 60$ K with an activation energy of 34 meV for Bi$_{89}$Sb$_{11}$ and 15 meV for Bi$_{85}$Sb$_{15}$. The inset shows the temperature dependence of d$\kappa_{zz}$ ($H_z$)/d$H_z$ between 4 and 8 T of TMZ sample #1 Bi$_{89}$Sb$_{11}$ and Bi$_{85}$Sb$_{15}$. (d) $\Delta\kappa_{zz}$ dependence on angle $\theta$ defined as $\theta = 0°$ for $H_\theta = H_z$ along [001], and $\theta = 90°$ for $H_\theta = H_y$ along [010].

| Sample name | Growth | x | Used for | density | mobility |
|---|---|---|---|---|---|
| | | at% | | cm$^{-3}$ | cm$^2$ V$^{-1}$ s$^{-1}$ |
| Sample #1 | TMZ | 10.5±0.5 | $\kappa_{zz}$ ($H_z$), $\kappa_{zz}$ ($H_y$) $\kappa_{zz}$ ($H_z$) Ag contacts | | |
| Hall | TMZ | 10.5±0.5 | Hall, resistivity | 3 × 10$^{15}$ (10 K) | 1.9 × 10$^6$ (10 K) |
| Sample #2 | Czochralski | 11.3±0.7 | Negative MR | | |
| | | 11.3±0.7 | Magneto-Seebeck | | |
| | | 11.3±0.7 | $\kappa_{zz}$ ($H_z$), $\kappa_{zz}$ ($H_y$) | | |
| Sample #3 | Czochralski | 11.3±0.7 | $\kappa_{zz}$ ($H_z$), $\kappa_{zz}$ ($H_y$) | | |
| | | 11.3±0.7 | $\kappa_{zz}$ ($H_z$)-angular dependence | | |
| Sample #4 | Czochralski | 11.3±0.7 | $\kappa_{zz}$ ($H_z$), $\kappa_{zz}$ ($H_y$) | | |
| Hall | Czochralski | 11.3±0.7 | Hall, resistivity | 1.4 × 10$^{16}$ (12 K) | 2 × 10$^4$ (12 K) |
| Sample #5 | TMZ | 15.1±0.7 | $\kappa_{zz}$ ($H_z$), $\kappa_{zz}$ ($H_y$) | | |
| Hall | TMZ | 15.1±0.7 | Hall, resistivity | 3 × 10$^{16}$ (10 K) | 4.5 × 10$^5$ (10 K) |
| Sample #6 | TMZ | 10.5±0.5 | $\kappa_{zz}$ ($H_z$), $\rho_{zz}$ ($H_y$) | as #1 | as #1 |
| semimetal | Bridgeman | 5±0.5 | $\kappa_{zz}$ ($H_z$) $\kappa_{zz}$ ($H_z$) Ag contacts, Hall, resistivity | 4.5 × 10$^{16}$ (79 K) | 8 × 10$^4$ (79 K) |

**Table 1 Samples used in this study.**